\documentclass[letter,12pt]{article}

\usepackage{cite}

\makeatletter
\renewcommand\@biblabel[1]{#1}
\makeatother

\usepackage{mathrsfs}

\usepackage{amssymb,amsfonts,amsmath}
\usepackage{setspace}
\usepackage{graphicx}
\usepackage{mathrsfs}
\usepackage{authblk} 
\usepackage[round,numbers]{natbib}
\usepackage[small]{caption}
\usepackage{fullpage}

\numberwithin{equation}{section}

\renewcommand*\theequation{\arabic{section}.\arabic{equation}}

\begin{document}

\title{\bf Genomic mutation rates that neutralize adaptive evolution and natural selection\vspace{50pt}}

\author[{\bf 1,2} \thanks{Author for correspondence (pgerrish@unm.edu) \\ Electronic supplementary material is available at \ldots}]{\bf Philip J. Gerrish}
\author[{\bf 3} ]{\bf Alexandre Colato}
\author[{\bf 4} ]{\bf Paul D. Sniegowski}
\affil[1]{\it Center for Evolutionary and Theoretical Immunology, Department of Biology, University of New Mexico, 230 Castetter Hall, MSC03-2020, 1 University of New Mexico, Albuquerque, New Mexico 87131 USA}
\affil[2]{\it Theoretical Biology \& Biophysics, Los Alamos National Laboratory, MS K710, Los Alamos, New Mexico 87545 USA}
\affil[3]{\it Departamento de Ci\^{e}ncias da Natureza, Matem\'{a}tica e Educa\c{c}\~{a}o, Federal University of S\~{a}o Carlos, UFSCar Araras, Brazil}
\affil[4]{\it Department of Biology, 213 Leidy Laboratories, University of Pennsylvania, Philadelphia, Pennsylvania 19104 USA}

\date{20 August 2012}
 
%% When adding keywords, separate each term with a straight line: | 

\begin{spacing}{1.5}

\maketitle

\newpage

\abstract{When mutation rates are low, natural selection remains effective, and increasing the mutation rate can give rise to an increase in adaptation rate. When mutation rates are high to begin with, however, increasing the mutation rate may have a detrimental effect because of the overwhelming presence of deleterious mutations. Indeed, if mutation rates are high enough: 1) adaptation rate can become negative despite the continued availability of adaptive and/or compensatory mutations, or 2) natural selection may be disabled because adaptive and/or compensatory mutations -- whether established or newly-arising -- are eroded by excessive mutation and decline in frequency. We apply these two criteria to a standard model of asexual adaptive evolution and derive mathematical expressions -- some new, some old in new guise -- delineating the mutation rates under which either adaptive evolution or natural selection is neutralized. The expressions are simple and require no \emph{a priori} knowledge of organism- and/or environment-specific parameters. Our discussion connects these results to each other and to previous theory, showing convergence or equivalence of the different results in most cases.\footnote{\it Journal of the Royal Society Interface}}

\vspace{20pt}
\begin{center}
\end{center}

\newpage
 
\section{INTRODUCTION}

\normalsize

Mutations are indiscriminate alterations of highly complex organisms and, as such, are much more likely to be harmful than beneficial. For an individual organism, therefore, an increase in the overall rate of mutation should be detrimental. In a population of organisms, however, natural selection disproportionately favors beneficial mutations and the net effect of increasing the overall mutation rate is thus less clear. 

%As mutation rate continues to increase, deleterious mutations can eventually be produced faster than they are selectively removed, thereby dragging down the fitness of the population. At high enough mutation rate, this effect can cause adaptation rate -- defined as rate of fitness increase -- to become negative.

\subsection{\textit{Previous studies}}

Generally speaking, the population-level effects of increasing the mutation rate have been studied separately under two artificial assumptions: the absence of beneficial mutations, and infinite population size. Only a handful of studies have relaxed both assumptions.

\subsubsection{\it Absence of beneficial mutations}

When beneficial mutations are assumed to be absent and population size is finite, fitness will undergo a slow but steady decline, because of the sluggish but largely irreversible accumulation of deleterious mutations. This process is especially pronounced in asexual populations, and it was in this context that the process was first described by Muller \cite{Muller:1932il} and later dubbed ``Muller's ratchet" \cite{Felsenstein:1974wq} and formalized by Haigh \cite{Haigh:1978wwa}. Under the relentless accumulation of deleterious mutations, fitness will decline monotonically. Most of the subsequent work on Muller's ratchet has focused on the rate of the ratchet, different factors affecting this rate, and in particular factors or conditions that can cause this rate to become negligible (i.e., that \emph{halt} the ratchet) \cite{Chao:1992vc,Green:1990vp,Andersson:1996uoa,Heller:1978ty,Loewe:2008iva,Gordo:2000tca,Kondrashov:1984twa,Fontanari:2003vd,Yuste:1999wb}. Increasing the genomic mutation rate can only accelerate Muller's ratchet.

\subsubsection{\it Infinite population size}

When population size is assumed to be infinite, populations whose adaptation is constrained -- i.e., populations in which beneficial mutations can occur but that have a maximum attainable fitness -- will eventually achieve an equilibrium fitness distribution shaped by the largely opposing forces of mutation and natural selection. Above a critical mutation rate dubbed the ``error threshold'' \cite{Eigen:2002vba,Eigen:1971tba}, this distribution becomes remarkably flat, indicating that a genotype's equilibrium frequency is essentially independent of its fitness. This conversion to a state of random fitness dispersion is reminiscent of a phase transition and, mathematically at least, the two are equivalent. The error threshold has been studied extensively under many different conditions that include recombination and departures from random mating \cite{Ochoa:1999tma,Nilsson:2000tgb,Boerlijst:1996uca}, viral complementation \cite{Sardanyes:2010boa}, and different static and dynamic fitness landscapes \cite{Takeuchi:2007tia,Nilsson:2000tgb,BoNHOEFFERfi:1992wua,Takeuchi:2005hma,Ochoa:1999tma,Gerrish:2009vw,Tejero:2010iba,Wiehe:1997vta}.

\subsubsection{\it Extinction}

The two classes of models described above -- Muller's ratchet and the error threshold -- encompass most previous characterizations of mutational degradation processes. Additionally, some work has superimposed demography onto both Muller's ratchet \cite{Lynch:1993tla,Gabriel:1993wy,Zeyl:2001uab} and error threshold \cite{Bagnoli:wk,Malarz:1998wo} models, finding a positive feedback between these processes and demographic decline toward extinction.

\subsubsection{\it Finite populations with beneficial mutations}

A few studies have addressed the effect of increasing the mutation rate when the two foregoing assumptions are relaxed -- i.e., when beneficial mutations are accounted for and populations are finite. Under these more realistic conditions, the fitness decline due to Muller's ratchet can be canceled out or even reversed by beneficial mutations, resulting in unchanging or increasing fitness. The effect of beneficial mutations on Muller's ratchet has been explored previously \cite{Goyal:2012gx,Bachtrog:2004tnb}; these studies focused on how the effects and relative fractions of beneficial vs. deleterious mutations would affect the adaptation rate and whether that rate was positive or negative. In the present study, we focus on how the genomic mutation rate affects the progress of adaptive evolution and the effectiveness of natural selection.

\subsection{\textit{Present study}}

\subsubsection{\it Neutralizing adaptive evolution}

When genomic mutation rate is low to begin with, an increase in this rate may be advantageous: the increased production of deleterious mutations can be of disproportionately small consequence because natural selection tends to eliminate deleterious mutations from the population, whereas the increased production of rare beneficial mutations can be of disproportionately large consequence because natural selection can cause the fixation of beneficial mutations from which the entire population benefits. Thus if a population's overall mutation rate is low to begin with, then an increase in the mutation rate can increase the rate at which beneficial mutations are fixed, thereby increasing the adaptation rate, where adaptation rate is defined as rate of increase in mean fitness. In other words, a positive correlation can exist between genomic mutation rate and adaptation rate. 

When genomic mutation rate is high to begin with, however, an increase in this rate may be disadvantageous because of excess deleterious mutation. While the consequence of deleterious mutations is still disproportionately small, it is less so at high mutation rates, because deleterious mutations can be produced faster than natural selection can remove them. At high mutation rates, therefore, a negative correlation can exist between genomic mutation rate and adaptation rate. 

The foregoing considerations indicate a non-monotonic relationship between mutation rate and adaptation rate, a relationship confirmed by simulation (Figs. \ref{fig:dwdt} and \ref{fig:timeavg}). In the present manuscript, we are interested in finding critical genomic mutation rates above which adaptation rate becomes negative. It seems reasonable to speculate that a negative adaptation rate, if sustained, would ultimately result in extinction.

\subsubsection{\it Neutralizing natural selection}

Evolution by natural selection proceeds through the appearance and subsequent fixation of adaptive and/or compensatory mutations. When mutation rate is low, virtually all adaptive and/or compensatory mutations produced have \emph{fixation potential}: all of them have the possibility, at least, of enduring the first few generations of random sampling (surviving genetic drift \cite{Haldane:1927ud}), outcompeting other adaptive and/or compensatory mutations (surviving the Hill-Robertson effect \cite{Hill:1966uua} or clonal interference \cite{Gerrish:1998uia}), and spreading to fixation. This is because, with low mutation rates, progress to fixation is relatively unhindered by deleterious mutations.

%When mutation rate is low, virtually all adaptive and/or compensatory mutations produced have \emph{fixation potential}: all of them have the possibility, at least, of enduring the first few generations of random sampling (surviving genetic drift), outcompeting other adaptive and/or compensatory mutations (surviving the Hill-Robertson effect or clonal interference), and spreading to fixation. This is because, with low mutation rates, progress to fixation is unhindered by deleterious mutations.

As mutation rate increases, however, the \emph{fixation potential} of adaptive and/or compensatory mutations is reduced: each such mutation founds a lineage whose growth is increasingly eroded by the accumulation of deleterious mutations. As mutation rate continues to increase, a point may be reached at which adaptive and/or compensatory mutations lose their fixation potential altogether, thereby neutralizing natural selection. We explore three particularly telling indicators that this point has been reached: 1) the fittest genotype in the population (e.g., an adaptive mutant) decreases in frequency, 2) the fittest genotype in the population has an equilibrium frequency (i.e., a mutation-selection balance frequency) very close to zero, and 3) a newly-arising fittest genotype is ultimately doomed to extinction with probability one.

\subsubsection{\it A key innovation: dynamical insufficiency}

In many of the previous investigations of mutational degradation processes, analogies are drawn to physical processes not least of which is the phase transition analogy. But the analogous physical processes typically occur on short time-scales during which the relevant parameters remain constant and convergence to equilibria occurs rapidly. This context affords the luxury of dynamically sufficient models and applicability of their steady-state analyses. In evolutionary biology, however, time scales are longer, relevant parameters cannot reliably be assumed to remain constant, and equilibria may rarely, if ever, be achieved. In the face of such long-term uncertainty, predictive accuracy seems unlikely; nevertheless, dynamically insufficient models may provide short-term predictive accuracy. Fisher's ``fundamental theorem of natural selection'' accurately predicts the evolution of fitness over the course of a single generation; by sacrificing dynamical sufficiency, this theorem achieves short-term predictive accuracy. Some of the conditions that we derive here (the more useful conditions) employ variations of this approach; they depend on statistical properties of the population that, by virtue of their \textit{intermediate} dynamical sufficiency, absorb contingencies and other surprises that are so characteristic of the biological world (see Discussion) and thereby may subsume many previous results that individually treat an array of different complexities and were derived under the purview of dynamical sufficiency.

\subsubsection{\it Our default application} 

The discovery and development of the error threshold sparked the imaginations of virologists, whose efforts to clear viral infections using antiviral drugs are bedeviled by the high mutation rates of many viruses. If mutation rate could be elevated even further through mutagenesis, error threshold theory suggested that viral populations might be driven extinct, thereby, in a sense, beating the virus at its own game \cite{Chen:2004wf,Chen:2005cj,Loeb:1999tca,Anderson:2004tu,Burger:1998ei}. Partly because of this historical context, we have adopted this particular application as our ``default'' application: unless otherwise stated, we have in mind the general aim of eradicating an unwanted population through mutagenesis and the inequalities we derive reflect this aim.

\subsubsection{\it Outline of the present study}

In the present study, we independently apply the two criteria described above to a standard, general model of fitness evolution in order to derive the conditions under which adaptive evolution and natural selection are neutralized. The conditions that we derive from criterion 1 range from sufficient to sufficient and necessary; however, it is the intermediate condition -- called ``sufficient and \emph{somewhat} necessary" -- that we believe is the most novel and perhaps the most practical. We apply criterion 2 both to a population in which the fittest genotype is resident (recovering the classical error threshold result in a new guise that lends itself to an alternative and perhaps more useful interpretation) and to one in which the fittest genotype is a newly-arising beneficial mutant.  

\section{RESULTS}

\subsection{\textit{The model}}\label{themodel}
We employ a standard model of adaptive evolution of an asexual population in which a genotype or class increases in log-frequency as the fitness of that genotype or class minus the mean fitness of the population. Mutation occurs among genotypes or classes as a diffusion process that is strongly biased in favor of deleterious mutations. Mathematical formulations of this model are given in the Appendix.

In what follows, we use the bracket notation in addition to the over-bar notation to denote expectation, and we use subscripted bracket notation to denote expectation with respect to the subscripted variable; for example, $\bar{x} = \left< x \right> $ denotes mean fitness, and $\left< \bar{x} \right>_t$ denotes mean fitness averaged over time. The models we employ are continuous in time and thus our measure of fitness $x$ corresponds to the log of fitness $w$ used in classical population genetics (discrete-time) models.

\subsection{\textit{Criterion 1: Adaptive evolution is neutralized}}
In this section, we employ a formulation of our model that is continuous in both time and fitness. We ask under what conditions adaptation will move backwards, i.e., under what conditions population mean fitness will decrease in spite of an inexhaustible supply of beneficial mutations.

\subsubsection{\it Sufficient and sufficient/necessary conditions} 

Adaptive evolution is neutralized when the long-term tendency of absolute fitness is to decrease, despite the availability of adaptive and/or compensatory mutations. A sufficient but not necessary version of this condition imposes $d\bar{x}/dt < 0$ at all times, where $\bar{x} = \left< x \right>$ is population mean fitness. The necessary and sufficient version of this condition is $\left< d\bar{x}/dt \right>_t < 0$. These conditions imply (Appendix) that adaptive evolution will be neutralized and fitness will in fact decline if the relation
\begin{equation}\label{condition1}
- U \left< \delta_x \right> \  >\  \sigma _{x}^{2}
\end{equation}
holds persistently (sufficient) or at least on average (sufficient and necessary), where $\sigma_x^2$ is variance in fitness, $U$ is genomic mutation rate, and $\delta_x$ is the effect of mutation on fitness (a random variable) and  $\left< \delta_x \right>$ is its average. (While this expression is given in terms of fitness, an equivalent expression is derived in terms of a fitness-related phenotype in the electronic supplemental material, or ESM.) If the effects of beneficial and deleterious mutations are considered separately, then $\left< \delta_x \right> ={{f}_{B}}{{m}_{B}}-{{f}_{D}}{{m}_{D}}$, where ${{f}_{D}}$ and ${{f}_{B}}$ are the fractions of all mutations that are deleterious and beneficial, respectively; ${{m}_{D}}$ and ${{m}_{B}}$ are the mean effects of deleterious and beneficial mutations on fitness, respectively. Biological considerations overwhelmingly support $f_D m_D \gg f_B m_B$, so the left-hand side of \eqref{condition1} will most likely be positive. By some estimates \cite{Perfeito:2007wa,Silander:2007gj,Desai:2007fh,Wloch:2001va,Joseph:2004wo,Sniegowski:2010be}, $f_B$ can be surprisingly high, however: 1) this does not necessarily imply high values of $f_B m_B$ \cite{Perfeito:2007wa}, and 2) it seems unlikely that $f_B m_B$ would ever exceed $f_D m_D$ simply because the ways to damage a highly complex entity (such as a living organism) far outnumber the ways to improve it. In the very unlikely case that $f_B m_B > f_D m_D$, condition \eqref{condition1} would present a contradiction, and fitness decline would be impossible regardless of $U$.

\textit{Critical mutation rate can be a moving target.} As evidenced by \eqref{condition1}, the critical mutation rate required to neutralize adaptive evolution is a function of the fitness variance. Increasing the mutation rate, however, will often cause a subsequent increase in fitness variance, in turn increasing the mutation rate required to satisfy \eqref{condition1}. In fact, classical population genetics (accounting for deleterious mutations only), and work by Rouzine \emph{et al.} \cite{Rouzine:2003en,Brunet:2008ff} and Goyal \emph{et al.} \cite{Goyal:2012gx} (accounting for beneficial and deleterious mutations) all indicate that, for low to moderate mutation rates, the fitness variance should tend toward $-U \left< \delta_x \right>$ following a perturbation in fitnesses and/or mutation rate. This suggests that an adjustment in the mutation rate (perhaps through increasing the dose of a mutagen, for example) to satisfy the condition $-U \left< \delta_x \right> > \sigma_x^2$ will be followed by an increase in fitness variance such that $\sigma_x^2 \rightarrow -U \left< \delta_x \right>$, thus necessitating a further increase in $U$ in order to 
maintain the relation $-U \left< \delta_x \right> > \sigma_x^2$. Frank and Slatkin \cite{Frank:2007vk} have pointed out that the tendency $\sigma_x^2 \rightarrow -U \left< \delta_x \right>$ represents mutation-selection balance (in fact, they mention this in the context of phenotypic evolution but the same notion applies). Figuratively, the condition $-U \left< \delta_x \right> > \sigma_x^2$ may be thought of as a mutation rate that persistently tips the balance in favor of mutation; alternatively, it may be thought of as a mutation rate persistently high enough to prevent convergence to mutation-selection balance. As $U$ is increased to maintain $-U \left< \delta_x \right> > \sigma_x^2$ in a continually adapting population, $\sigma_x^2$ will eventually reach a maximal value (due to finite population size) and, at this point, the value of $U$ need not increase further to satisfy $-U \left< \delta_x \right> > \sigma_{max}^2$. In Fig. \ref{fig:realtime}\emph{B}, the genetic variance in fitness is measured in simulated populations every 100 generations and $-U \left< \delta_x \right>$ is set at ten percent above $\sigma_x^2$, thereby maintaining $-U \left< \delta_x \right> > \sigma_x^2$. For a long time, the positive feedback 
between mutation rate and fitness variance results in escalating adjustments to the mutation rate; after some time, however, the variance appears to achieve a maximum so that the mutation rate required for continued fitness decline levels off.

\subsubsection{\it Sufficient and \emph{somewhat} necessary condition} 
 
So far, we have derived conditions that lie at opposite ends of the spectrum from sufficiency to sufficiency-and-necessity. From a practical standpoint, however, both are of limited utility. Condition \eqref{condition1} insures declining fitness only for the current generation. The sufficient condition is that this relation hold persistently, but this condition may be frustratingly elusive because it fails to anticipate the change in fitness variance that typically follows an adjustment to the mutation rate. For this condition to be enforced in practice, therefore, frequent measurements of $\sigma_x^2$ would be required, followed by adjustments in $U$ (for example, by increasing the dose of a mutagen), if needed, to maintain the relation \eqref{condition1} (as in Fig. \ref{fig:realtime}\emph{A,B}). In practice, therefore, the sufficient condition amounts to a rather inconvenient protocol. The sufficient-and-necessary condition, that \eqref{condition1} hold on average, requires long-term future knowledge of population fitnesses that is generally not attainable in practice. Here, we derive conditions that lie somewhere in the middle of the spectrum from sufficiency to sufficiency-and-necessity and that have increased practical applicability.

To this end, we temper our sufficient and necessary condition: instead of requiring that the \emph{long-term} average gradient oppose selection, we now require only that the \emph{medium-term} average gradient oppose selection. We will denote this intermediate condition as $\left< d\bar{x}/dt \right>_t ^{r} < 0$, where $r$ denotes the number of future generations over which to take the average. In order to enforce this condition, however, one needs a way to predict the near-future course of evolution; an algorithm for doing this is outlined in \cite{Gerrish:2012bo}. There, it is shown that prediction of the near-future course of evolution can be achieved by a time-discretization of a hierarchy of cumulant equations. 

Using the equations for fitness evolution derived in \cite{Gerrish:2012bo} and imposing $\left< d\bar{x}/dt \right>_t ^{r} < 0$, the condition under which adaptive evolution is neutralized may be written:
\begin{equation}\label{fitness2}
- U \left< \delta_x \right> > \frac{1}{r} \sum\limits_{\tau = 0}^{r-1}\kappa_2 (\tau),
\end{equation}
where the future fitness variances (or second cumulants), $\kappa_2 (\tau) = \sigma_x^2(\tau)$, are computed from the set of recursions $\kappa_{i}(\tau+1) = \kappa_{i}(\tau)+\kappa_{i+1}(\tau) + U m_{i}$ as outlined in \cite{Gerrish:2012bo} (also, see Appendix); $\kappa_{i}(\tau)$ denotes the $i^{th}$ cumulant at generation $\tau$; $\tau=0$ denotes the present generation (called ``now''), $\tau =1$ denotes one generation from now, $\tau = 2$ denotes two generations from now, etc., and $r$ is the ``predictive reach", i.e., $r$ is how many generations into the future the algorithm can be trusted to predict. An alternative condition that errs conservatively is: $- U \left< \delta_x \right> > \operatorname{max}(\kappa_2 (0),\kappa_2 (1),...,\kappa_2 (r))$. (See ESM for equivalent phenotypic expressions.) The appearance of these equations is deceptively simple because as $U$ is changed, the predictions for $\kappa_2 (\tau)$ will change, i.e., the equations look explicit when in fact they are implicit for $U$. (They are implicit for $U$ because a certain degree of circularity is required by their intermediate dynamical sufficiency, which anticipates future changes in $\sigma_x^2$ without requiring knowledge of organismal and environmental parameters; in practice, this fact only imposes the slight inconvenience of having to use an iterative procedure in the calculations.) 

\subsection{\textit{Criterion 2: Natural selection is neutralized}}
The approach that derives from this criterion takes its lead from statistical physics, where an ``order parameter" quantifies the degree of order present in the system at hand. Order in an evolving population is brought about through the action of natural selection on genetic variation. In evolution, a natural choice for an order parameter is the frequency of the fittest genotype. If natural selection is operational, the fittest genotype should persist at reasonable frequency despite recurrent mutation away from this genotype, and this frequency is thus indicative of the amount of order present in the population. As mutation rate increases, the frequency of the fittest genotype will decrease, indicating a decrease in the overall order present. At a sufficiently high mutation rate, the amount of order will approach zero.

\subsubsection{\it Sufficient condition}

Here, we have in mind a population that is heterogeneous and that is predominated by a fittest genotype whose frequency is $u_0$. Our sufficient condition is derived by finding the mutation rate that causes the frequency of the fittest genotype to decrease relative to its mutational neighbors: $du_0/dt < 0$, persistently. Solving for the mutation rate that insures this inequality gives rise to the condition: 
\begin{equation}\label{ET0}
U f_D > (x_0 - \tilde{x})(1/(1-u_0)-1/L u_0)^{-1},
\end{equation}
where $L$ is the size of the deleterious genome, $x_0$ is the fitness of the fittest genotype; $\tilde{x} = \sum\limits_{j=1}^{L} x_j \hat{u}_j$, and $\bar{x} = \sum\limits_{j=0}^{L} x_j \hat{u}_j$, from which we have the useful relation $(1-u_0)(x_0 - \tilde{x}) = x_0 - \bar{x}$. In a finite population, $x_0$ is the maximum fitness found in the population, and $\tilde{x}$ is the average fitness of everybody else: $\tilde{x} = \frac{1}{\#S} \sum_{i \in S} x_i$, where $S$ is the subset of the population that has fitness less than the maximum and $\#S$ is the number of individuals in that subset.. If it is the case that the population is finite and $L \gg N$, then the term $1/L u_0 \le N/L\approx 0$, giving rise to the condition: 
$
U f_D \gtrsim (1-u_0)(x_0 - \tilde{x}) = x_0 - \bar{x}
$
(reported in Table 1). In our simulations, we assume an infinite genome ($L \rightarrow \infty$) and finite population size; under these conditions, 
$
U f_D > x_0 - \bar{x}
$ is exact.
In a continually adapting population, $u_0$ will be small most of the time, in which case this expression may be used interchangeably with:
$
U f_D \gtrsim x_0 - \tilde{x}
$
(employed in Fig. \ref{fig:realtime}).

\subsubsection{\it Sufficient and necessary conditions}
(1) \textit{Mutational degradation of an established fittest genotype.}
Here, we have in mind a population that is heterogeneous but that has been predominated by a fittest lineage for some time. To determine the amount of order in this population, we compute its order parameter, $\hat{u}_0$: the equilibrium frequency of this fittest lineage relative to its mutational neighbors (genotypes that differ from the fittest lineage by mutation). We are especially interested in what happens to the order parameter as mutation rate increases. 

Analysis of the evolutionary model at equilibrium reveals that indeed the order parameter $\hat{u}_0$ decreases with increasing mutation rate (Appendix). The approach of $\hat{u}_0$ toward zero as $U$ increases is characterized by an inflection point that becomes increasingly sharp as deleterious genome size $L$ increases. The mutation rate at which the inflection point occurs is found by solving for the critical mutation rate $U_c$ that satisfies $\partial^3 \hat{u}_0 / \partial U^3 = 0$. As $L$ increases, 
$
U_c f_D = \mu_c L  \rightarrow x_0 - \tilde{x},
$
where $f_D$ is again the fraction of mutations that are deleterious, and $\mu_c$ is the critical point mutation rate. From this result, natural selection may reasonably be expected to be neutralized when mutation exceeds the critical rate:
\begin{equation}\label{ET}
U f_D \ge x_0 - \tilde{x}.
\end{equation} 
This is the classical ``error threshold" result in new guise. It is an equilibrium result and its practical use would therefore require knowledge of long-term future states of the population.

The equilibrium frequency of the fittest class or genotype at the ``error threshold", while greatly reduced, is still greater than the frequencies of neighboring genotypes: at $\mu = \mu_c$, the fittest genotype has frequency
$
\hat{u}_0 \approx \frac{1}{L} + \sqrt{\frac{1}{L}}.
$,
whereas mutational neighbors have frequency $\hat{u}_i < \frac{1}{L}$.
This stands in contrast to common notions about the error threshold as creating a competitive reversal that leads to the subordination and/or loss of the fittest genotype. In a finite population, the fittest genotype will be deterministically lost from the population at the error threshold only if $N \lesssim \sqrt{L}$. To put this condition in perspective, we consider a strain of \emph{Escherichia coli} that has a genome of length $L \approx 4.6 \times 10^6$ base pairs; if we make the very conservative assumption that mutation at any position on the genome will affect fitness, then any population larger than $\sqrt{L} \approx 2145$ will deterministically retain the fittest genotype at the error threshold. 

Despite the persistence and continued dominance of the fittest genotype, the error threshold nevertheless marks a point at which the frequencies of the different genotypes are so severely eroded by mutation that their frequencies are clearly not indicative of their fitness. This disabling of natural selection is apparent in the relation:
$
\operatorname{cov}(u,x) |_{\mu = \mu_c} \approx \bar{s}/L,
$
where $\bar{s} \approx - \left< \delta_x \right>$. For large genomes, therefore, the covariance between fitness and frequency -- an indicator of the efficacy of natural selection -- is very small at the error threshold (but still positive). Additionally, the extent to which natural selection has become ineffective is reflected by the amount of \emph{dis}order present in the equilibrium population; a standard index of disorder is the Shannon entropy which, at the error threshold, is approximately equal to $\operatorname{log}_2 L$.

(2) \textit{Mutational degradation of a newly-arising fittest genotype.}
Here, we have in mind an asexual population that is heterogeneous and in which a beneficial mutation emerges. This mutation creates a newly-arising ``fittest genotype" whose subsequent growth depends on the persistence of that genotype within the growing lineage, despite recurrent mutation away from that genotype. The newly arising fittest genotype has fitness $x_0$, and the rest of the population has average fitness $\tilde{x}$, as before. In a single generation, the new lineage grows by a factor $R = e^{x_0 - \bar{x}} = e^{(x_0 - \tilde{x})(1-u_0)}$. Accumulation of deleterious mutations occurs most rapidly early in the growth of a lineage \cite{Fontanari:2003vd}, when $u_0 \approx 0$, suggesting the approximation $R \approx e^{x_0 - \tilde{x}}$. Previous studies show that genomic mutation rates that cause the degradation of the newly-arising fittest genotype 
must satisfy $U f_D \ge \operatorname{log} R$  \cite{Fontanari:2003vd,Bull:2008gw,Bull:2007cda}. The extinction of a newly-arising fittest genotype is therefore predicted to occur when:  
\begin{equation}\label{colato}
U f_D  \gtrsim x_0 - \tilde{x}
\end{equation}
(Compare to \eqref{ET}.) This result was originally derived for an independent asexual population growing without bound at discrete-time rate $R$ \cite{Fontanari:2003vd} and was later re-derived in a way that more explicitly allowed for purifying selection and dubbed the ``lethal mutagenesis" threshold \cite{Bull:2007cda,Bull:2008gw} for unboundedly growing viral and bacterial populations. This result should also apply, however, to lineages growing within a population as a consequence of positive \emph{relative} fitness ($x_0 - \tilde{x}>0$). Finite population size restricts applicability to lineages that begin to decline in frequency before being affected by population size constraints, which seems likely to account for many such lineages when at or near the critical mutation rate (but see \cite{Gerrish:2007cv}). Those lineages that do achieve higher frequencies are likely to become fixed in the population, in which case the relevant condition was derived in the previous subsection: $U f_D \ge x_0 - \tilde{x}$ (condition \eqref{ET}). It thus seems reasonable 
to conjecture that whatever the maximum frequency achieved by the new lineage, the condition is well approximated by \eqref{colato}.

\section{DISCUSSION}

\subsection{\textit{Practical use of the equations}}

\textit{Why are accurate predictions desireable?} On the surface, it seems that if one has the ability to increase mutation rate, perhaps through the use of a chemical mutagen, then to drive a population extinct, one needs only to increase the mutation rate by a large amount, perhaps by administering a high dose of mutagen. The problem with this approach is that, in real populations, variation in mutation rate is inevitable and resistance to a mutagen can appear. A large increase in the mutation rate can create strong selection pressure for a lowered mutation rate, and a reduction in the mutation rate may thus evolve in short order. Our own work on this (in prep.) has shown that \emph{Escherichia coli} evolves resistance to a nucleoside analogue mutagen, administered at a fairly high dose, in just a few generations. If one could increase the mutation rate to a level that is high enough to cause extinction, but not too high, selection for resistance could in principle be reduced considerably and the evolution of resistance 
might be prevented. Accurate predictions for the critical mutation rate required for extinction may therefore aid in the practical implementation of chemical mutagenesis, and the evolution of resistance might be prevented. Indeed, our equations and simulations would suggest an improved protocol in which a mutagen is administered in incrementally increasing dose (reflected in Fig. \ref{fig:realtime}). 

\textit{Timeframe of applicability.} The equations derived here are similar in their generality and robustness; however, they differ amongst themselves in one aspect of practical relevance, namely, their timeframe of applicability. Under criterion 1, this timeframe ranges from short-term (sufficient) to medium-term (sufficient and somewhat necessary) to long-term (sufficient and necessary). Under criterion 2, the timeframe is short-term (sufficient) or long-term (sufficient and necessary). The long-term results might potentially be applied approximately using a running-average approach that is necessarily somewhat arbitrary, but technically correct application of these results requires information about long-term future states of the population that would not be obtainable in practice. When the mutation rate is adjusted according to fitness measurements from a population taken in real time, the correct equations to use are the short-term and medium-term conditions. These conditions are applied in simulation studies of which representative runs are presented in Fig. \ref{fig:realtime}; 
there, adjustments to the mutation rate are made in real time and the short-term and medium-term conditions derived under criterion 1 (labeled ``variance'' and ``variance-projection'' thresholds, respectively) perform well, whereas the conditions derived under criterion 2 (error threshold) appear to be less well-suited to such real-time application.

\textit{Adaptation in a static environment.} A population adapting in a static environment typically has a limited, non-renewable supply of available beneficial mutations (barring intransitive interactions). As the population adapts, therefore, the supply of available beneficial mutations is slowly depleted; as a consequence, mean fitness may increase and subsequently decrease, and fitness variance may also change over time, thereby changing the minimal mutation rate prescribed by \eqref{condition1}. This is shown schematically in Fig. \ref{fig:schematic}: in static environments and, generally speaking, in environments where the supply of beneficial mutations can change over time, adaptive evolution or natural selection may be neutralized not as a result of changes in the mutation rate (i.e., changes in $-U \left< \delta_x \right>$) but as a result of changes in the requirements on the mutation rate (i.e., changes in $\sigma_x^2$). This is the basis of an illustrative application to immunology that is described in the ESM.

\subsection{\textit{Connections to previous theory}}

\textit{Fisher and Kimura.}
Fisher's ``fundamental theorem of natural selection" states that, when $x$ is defined as additive genetic fitness, $d\bar{x}/dt = \sigma_x^2$ quite generally \cite{Frank:1992ug,Fisher:1930wya}. Fisher's theorem shows that this particular component of fitness can only increase (variance is a non-negative quantity); consequently, this component of fitness has accurately been called the ``adaptive engine" of natural selection \cite{Grafen:2003tj}. This component of fitness, however, must be conserved over time, for example, in the transmission from parent to offspring for Fisher's theorem to apply. If there is a component of fitness that is not conserved, then to find the change in \emph{total} mean fitness of the population (conserved and non-conserved), one must use the chain rule: $d\bar{x}/dt = \int \left( \frac{dx}{dt} u(x,t) + x \frac{d}{dt} u(x,t) \right) dx$ which, together with $\frac{d}{dt} u(x,t) \equiv (x - \bar{x})u(x,t)$, yields $d\bar{x}/dt = \sigma_x^2 + \left< dx/dt \right>_x$ -- a 
fact pointed out by Kimura \cite{Kimura:1958vr}. (Note the subscript $x$, indicating that here the average is taken over fitness or, equivalently, over individuals in the population.) As a general rule, $\left< dx/dt \right>_x$ will be negative, because random alterations in the organism or its environment are more likely to decrease the organism's fitness than to increase it. This simple calculation illustrates the fact that Fisher's fundamental theorem applies only to that subset of the population whose additive genic fitness is conserved over the period of time in question: only for this particular subset of the population are we guaranteed that mean fitness will not decrease.

Applying our criterion 1 to Kimura's equation yields a more general condition for the disabling of adaptive evolution:
\begin{equation}\label{kimura}
- \left< dx/dt \right>_x > \sigma_x^2
\end{equation}
must hold persistently or at least on average. Here, the mechanism of change in individual fitnesses over time is not specified. If we specify that the mechanism of change is mutation, then $\left< dx/dt \right>_x = U \int_{\delta_x} \delta_x g( \delta_x,t) = U \left< \delta_x \right>$, where $g(\delta_x,t)$ is the distribution of mutational effects on fitness, and we recover Eq. \eqref{condition1}.

%\begin{figure}[h!]
%\begin{center}
%\scalebox{.15}{\includegraphics{fig1.jpg}}
%\caption{\label{fig:epistasis}}
%\end{center}
%\end{figure}

\textit{The error threshold.} As previously stated, Eqs. \eqref{ET} and \eqref{colato} are the error threshold in new guise. The original work on the error threshold due to Eigen \cite{Eigen:1971tba} derives a minimum value for the ``quality factor" -- the probability of complete fidelity of replication -- that is needed to maintain the efficacy of natural selection and thus to support life. This minimum value is given by $\mathscr{Q}_{min} = (\bar{\mathscr{A}}_{k \ne m} + \mathscr{D}_{m} - \bar{\mathscr{D}}_{k \ne m})/\mathscr{A}_{m}$ (Eq. II-45 in \cite{Eigen:1971tba}), where $\mathscr{A}_{m}$ and $\mathscr{D}_{m}$ are the birth and death rates of the fittest genotype (the ``master sequence"), respectively, $\bar{\mathscr{A}}_{k \ne m}$ and $\bar{\mathscr{D}}_{k \ne m}$ are the mean birth and death rates of the rest of the population (individuals that do not carry the ``master sequence"). The quantity $\mathscr{A}_m (1 - \mathscr{Q}_{min})$ is the expected number of deleterious mutants produced by a single 
replication event of the fittest genotype. This quantity, in our notation, is $U f_D$; furthermore, $\mathscr{A}_{m} - \mathscr{D}_{m}$ is equivalent to our $x_0$ and $\bar{\mathscr{A}}_{k \ne m} - \bar{\mathscr{D}}_{k \ne m}$ is equivalent to our $\tilde{x}$. Eigen's result may thus be rewritten in our notation as requiring $Uf_D < x_0 - \tilde{x}$ for the effectiveness of natural selection to be maintained, or conversely, $Uf_D \ge x_0 - \tilde{x}$ for natural selection to be neutralized.

In work subsequent to Eigen's original publication, the varied presentations of his error threshold result are usually rearrangements of this simple expression: $q_{min} ^L = \sigma^{-1}$, where $q_{min}$ is the minimum per-nucleotide replication fidelity required for survival ($q_{min} = 1 - \mu_c$), $L$ is the length of the deleterious genome, and $\sigma$ is the ``superiority parameter", defined as $\sigma = 1/\mathscr{Q}_{min}$. Rewriting reveals an interesting biological requirement: $\log{q_{min}^L} \approx -\mu_c L$ gives rise to the relation: 
\begin{equation}\label{ETx}
\mu_c \approx (\mbox{something})/L. 
\end{equation}
This inverse relation between $\mu_c$ and $L$ intrigued its discoverers to the extent that the ``something'' was all but ignored. It was since discovered, however, that observations of $\mu L$ are surprisingly constant across microbial taxa \cite{Drake:1991tj,Drake:1999vl} (indeed, it has been conjectured that this is the case precisely because of the inverse relation between $\mu_c$ and $L$). The relative constancy of $\mu L$ across taxa suggests that the ``something'' may in fact be quite relevant to the fate of a population; furthermore, $L$ will probably not change on time scales pertinent to extinction-by-mutation. These considerations shift the focus to $\sigma$. Its name together with its traditional presentation obfuscates the fact that $\sigma$ is a population-dependent \emph{quantity} and not an organism-dependent \emph{parameter}. Our new presentation of this old result shifts the emphasis from critical \emph{point} mutation rate $\mu_c$ vs. genome length $L$ to critical \emph{genomic} mutation rate $\mu_c L$ vs. the myriad biological, ecological and environmental factors that are not explicitly part of the equation but that are absorbed by the quantity $\sigma$ or, in our formulation, $x_0 - \tilde{x}$.

\subsection{\textit{Connections among results presented here}}

Our first criterion is the sustained decline of absolute fitness, whereas our second criterion is the inefficacy of natural selection. We now show that, despite these perhaps disparate criteria, the resulting conditions for extinction connect through classical population genetics. Criterion 1 gives rise to the condition $-U \left< \delta_x \right> > \sigma_x ^2$, and criterion 2 gives rise to the condition $U f_D \ge x_0 - \tilde{x}$. Our comparison of these two results proceeds by multiplying both sides of the second condition by $m_D$ to obtain $U f_D m_D \ge (x_0 - \tilde{x}) m_D$. First, we note that the left-hand side is $U f_D m_D \approx -U \left< \delta_x \right>$ because most mutations are deleterious. Next, we focus on the right hand side of the inequality. As a population approaches the error threshold (i.e., as this inequality approaches equality), the size of the fittest class approaches zero and it is the case that $\frac{1}{\#S} \sum_{i \in S} x_i \rightarrow \frac{1}{
N} \sum\limits_{i=1}^{N} x_i$, or $\tilde{x} \rightarrow \bar{x}$. The quantity $x_0 - \bar{x}$ is known in classical population genetics as the genetic load, and it is known to converge to the deleterious mutation rate $U f_D$. Furthermore, it is known that if mutations are assumed to have a fixed deleterious effect, $m_D$, then the number of accumulated mutations becomes Poisson distributed with mean $U f_D / m_D $ \cite{Haigh:1978wwa}. The variance in number of accumulated mutations is the same as the mean, and the variance in \emph{fitness} is therefore $\sigma_x^2 = \left( U f_D / m_D \right) m_D^2 = U f_D m_D$. As the error threshold is approached, therefore, the right-hand side becomes $(x_0 - \tilde{x}) m_D \rightarrow (x_0 - \bar{x}) m_D = U f_D m_D = \sigma_x^2$. 

\subsection{\textit{Borrowed robustness}} 

Fisher's fundamental theorem of natural selection is known to be extraordinarily accurate in spite of numerous complexities that are characteristic of real populations. Because \eqref{condition1} is implicit in the results of Fisher and Kimura, therefore, we expect these results to be quite robust to numerous biological complexities. Furthermore, the convergence we have demonstrated between \eqref{condition1}, \eqref{ET0} and \eqref{colato} lead us to believe that the classical error threshold result is similarly robust, although it does not appear to perform as well in real time (Fig. \ref{fig:realtime}). Figure \ref{fig:timeavg} together with the plots we have posted in the ESM -- and many others not posted -- demonstrate the robustness of \eqref{condition1} (and by inference \eqref{fitness2}) to a wide range of complexities, including finite genome effects, the effects of finite population size (including Muller's ratchet), epistatic interactions among mutations, environmental noise (random changes in fitness caused by unspecified factors), 
an evolving mutational robustness modifier, compensatory mutations whose rate increases with decreasing fitness, an evolving mutation rate modifier, and a fraction of mutations that are lethal.\\

\small{
Special thanks to Cristian Batista for insightful explanations of the error threshold and to Isabel Gordo for helping make connections among the different theories. We also thank Michael Lassig, Paul Joyce, Nico Stollenwerk, Gabriela Gomes, Ana Margarida Sousa, Jorge Carneiro and Josep Sardany\'{e}s for helpful discussions. Much of this research was developed thanks to fertile environments provided by two institutes: the Kavli Institute for Theoretical Physics in Santa Barbara, CA (2011 Microbial and Viral Evolution workshop), and the Instituto Gulbenkian de Ciências in Oeiras, Portugal. This work was supported by the US National Institutes of Health grants R01 GM079843-­‐01 (PJG/PDS) and ARRA PDS\#35063 (PJG/PDS), and European Commission grant FP7 231807 (PJG).}

%\bibliographystyle{authoryear}
%\bibliography{authoryear}
 
%% == end of paper: 
 
%% Optional Materials and Methods Section 
%% The Materials and Methods section header will be added automatically. 
 
%% Enter any subheads and the Materials and Methods text below.
\appendix 
\section*{APPENDIX}
\renewcommand*\theequation{A.\arabic{equation}}
%\sffamily 
%\small
The results we describe in the main text derive from two manifestations of a standard model of evolution described verbally under the subheading ``The model". Here we give the mathematical details of those manifestations:

\textbf{Model in continuous fitness for Criterion 1.} We let $u(x,t)$ denote the density of individuals in the population with log-fitness $x$ at time $t$. Mutation can create "jumps" in log-fitness whose size has probability density $g(\phi,t)$ at time $t$. Under selection and mutation, a population's evolution is described by:
\begin{equation}\label{second}
\begin{split}   \frac{\partial }{\partial t} u(x,t)\ &=\ (x-\bar{x})\ u(x,t)\ \\
&+\ U \left( \int\limits_{-\infty }^{+\infty }{u(x-\phi ,t)g(\phi ,t)d\phi \ -\ u(x,t)} \right),
\end{split} \end{equation}
where $\bar{x} = \int_{-\infty}^{+\infty} x u(x,t) dx$. If we apply the standard diffusion approximation to the mutation term, this equation becomes 
\begin{equation}\label{approx1}
\frac{\partial}{\partial t} u(x,t) = (x - \bar{x}) u(x,t) + U \mbox{\textbf{M}} u(x,t).
\end{equation}
Mutation operator, $\mbox{\textbf{M}} = D_x \frac{\partial^2}{\partial x^2} - d_x \frac{\partial}{\partial x}$, where $D_x =  \frac{1}{2} f_B (m_B^2 + \sigma_B^2) + \frac{1}{2} f_D (m_D^2 + \sigma_D^2)$ and $d_x = f_B m_B - f_D m_D$; $f_B$ is the fraction of all mutations that are beneficial (``beneficial fraction"), $f_D$ is the deleterious fraction; $m_B$ and $m_D$ are the mean effects of beneficial and deleterious mutations on fitness, respectively; $\sigma_B^2$ and $\sigma_D^2$ are the variances in those effects. We multiply both sides of \eqref{approx1} by $x$ and integrate over all $x$ to obtain $\dot{\bar{x}} = \sigma_x^2 + U \int_{-\infty}^{+\infty} \mbox{\textbf{M}} x u(x,t)dx$. Under the reasonable assumption that $u(x,t)$ has compact support in $x$, integration by parts gives $\dot{\bar{x}} = \sigma_x^2 + U \left< \delta_x \right>$, where $\left< \delta_x \right> = f_B m_B - f_D m_D$. The condition $\dot{\bar{x}} < 0$ reflects the disabling of adaptive evolution and is met when $-U \left< \delta_x \right> > \sigma_x^2$.

\textbf{Model in discrete fitness for Criterion 2.} As an indication of the amount of order in the system at hand, we would like to know the frequency of the fittest genotype relative to its mutational neighbors. The dynamics of this genotype and its mutational neighbors (genotypes that differ from the fittest genotype by mutation) are given by this set of equations: 
\begin{equation}\label{ET1}
{{\dot{u}}_{i}}=({{x}_{i}}-\sum\limits_{j=0}^{L}{{{x}_{j}}{{u}_{j}}}){{u}_{i}}-L\mu {{u}_{i}}+\mu \sum\limits_{j\ne i}^{{}}{{{u}_{j}}},
\end{equation}
where 		${{u}_{0}}$ = the frequency of the fittest genotype (the order parameter),
${{u}_{i}}$ = the frequency of mutational neighbor, $i=1,2,3,...,L$, 
		${{x}_{i}}$ = fitness of genotype $i$, and
		$\mu $ = point mutation rate.

The equation for the fittest genotype $u_0$ may be written as: 
\begin{equation}\label{ET1a}
 \dot{u}_0 = (x_0 - \tilde{x})(1- u_0) u_0 - \mu L u_0 + \mu(1-u_0),
\end{equation}
where $x_0$ is the fitness of the fittest genotype and $\tilde{x}$ is the average fitness of everybody else: $\tilde{x} = \sum\limits_{j=1}^{L} x_j u_j / (1-u_0)$.
We noted that $x_0 - \tilde{x}$ is not relative fitness; a possible interpretation of the value $x_0 - \tilde{x}$ is that it is the reproductive ``payoff" in a game played by the fittest genotype against everybody else. 

To find the ``sufficient condition" for neutralizing natural selection, we solve for the $\mu$ that satisfies $\dot{u}_0 < 0$ and then make the substition $\mu L = U f_D$, giving the condition:
\begin{equation}\label{ET0a}
U f_D > (x_0 - \bar{x}) \left( 1 - \frac{1-u_0}{L u_0} \right)^{-1},
\end{equation}
where $x_0 - \bar{x} = (x_0 - \tilde{x})(1-u_0)$ and $\bar{x}$ is simply the mean fitness of the population. In a continually adapting population, $u_0$ will be small most of the time, and $\bar{x} \approx \tilde{x}$. Thus, for large $L$ and $L \gg N$, \eqref{ET0a} may be approximated by $U f_D \gtrsim x_0 - \tilde{x}$.

To find the ``sufficient and necessary conditions", we derive the order parameter $\hat{u}_0$, the equilibrium value of $u_0$ that satisfies $\dot{u}_0 = 0$. Analysis of this equilibrium reveals that indeed the order parameter $\hat{u}_0$ decreases with increasing mutation rate. The approach of $\hat{u}_0$ has a sharp inflection point whose position is found by setting $\partial^3 \hat{u}_0 / \partial \mu^3 = 0$; for large genome size $L$, it occurs at a critical mutation rate given by
$
U_c f_D = \mu_c L = x_0 - \tilde{x},
$
where $U_c$ is the critical genomic mutation rate, $f_D$ is the deleterious fraction, $\mu_c$ is the critical point mutation rate. In a finite population, $x_0$ is the maximum fitness found in the population, and $\tilde{x} = \frac{1}{N} \sum_{i \in S} x_i$, where $N$ is population size and $S$ is the subset of the population that has fitness less than the maximum.  This is the error-threshold result \cite{Eigen:1971tba,Eigen:2002vba} in new guise and may be phrased as follows: natural selection is neutralized when:
\begin{equation}\label{ET2}
U f_D \ge x_0 - \tilde{x}.
\end{equation} 

\textbf{Calculating the ``sufficient and somewhat necessary" conditions under Criterion 1.} To compute the ``sufficient and somewhat necessary" conditions requires projection of cumulants $\kappa_i(\tau)$ over a period of $r$ generations into the future. Recurrence relations that do this are developed in \cite{Gerrish:2012bo}:

The terms of the sum in \eqref{fitness2} are computed from the recurrence relation: ${{\kappa }_{i}}(\tau )~~=\ \ {{\kappa }_{i}}(\tau -1)~+\ {{\kappa }_{i+1}}(\tau -1) + U {{m}_{i}}$ for all $i\ge 1$, where $\kappa_{i}(\tau)$ is the $i^{th}$ cumulant in fitness at a time $\tau$ generations from now, $U$ is genomic mutation rate and $m_{i}$ is the $i^{th}$ raw moment of the distribution of mutational effects on fitness.  

The practical implementation of condition \eqref{fitness2} requires some care. The procedure outlined in \cite{Gerrish:2012bo} provides methods for estimating the $m_j$. These parameters cannot be estimated separately from $U$; only their products $U m_j$ can be estimated, if the equations are left in non-parametric form. The obvious remedy is to make the equations parametric by writing the known expressions for the moments of an assumed distribution in place of $m_j$. Then, the parameters to be estimated are $U$ and the limited number of parameters of the assumed distribution, and $U$ can then be estimated separately. If one's objective is to monitor a population's risk of extinction, or to drive a population extinct through mutagenesis, however, a less obvious remedy may apply. In such cases, absolute mutation rates may be irrelevant, and the effects of an increased (or decreased) mutation rate can be predicted by simply multiplying the 
estimates of $U m_j$ by the factor by which mutation rate is increased (or decreased). In such cases, therefore, the equations may be left in non-parametric form.

% Bilbiography Style

%\bibpunct{}{}{,}{n}{,}{,}
\bibliographystyle{rspublicnat2}
\bibliography{pjg5a}

\newpage

\subsection*{Figure Captions} 

Figure 1. Time-averaged fitness gradients from simulations as a function of genomic mutation rate. Simulations are fully stochastic and individual-based; populations are asexual. Population size is 10,000; fraction of mutations that are deleterious is constant at 0.1; fraction of mutations that are beneficial is constant at 0 (green), $10^{-5}$ (orange), $10^{-4}$ (blue), and $10^{-3}$ (red); effects of mutations are drawn at random from an exponential distribution with mean 0.03. At high enough mutation rates, the rate of fitness increase becomes negative (indicating persistent fitness decline), from which inference of eventual extinction seems reasonable.\\

\noindent Figure 2. (\emph{A}) Time-averaged adaptation rate as a function of genomic mutation rate. The point at which adaptation rate becomes negative marks the threshold mutation rate, indicated by the blue vertical line. (\emph{B}) Predictions for the threshold mutation rate. Yellow triangles plot the variance threshold given by Eq. \eqref{condition1}; blue diamonds plot the error threshold given by Eq. \eqref{ET}; red line plots genomic mutation rate $U$. Where threshold predictions intersect with the red line marks the predicted threshold mutation rate for these simulations, and coincides exactly with the observed threshold mutation rate in panel (\emph{A}). Each point represents an average taken over the full time course of eight fully stochastic, individual-based simulations of evolving asexual populations. Population size was 10,000, fraction of mutations that were beneficial and deleterious were 0.001 and 0.5, respectively. Ten percent of deleterious mutations were lethal; otherwise, beneficial and 
deleterious mutations were drawn from an exponential distribution with mean 0.03. Epistasis among deleterious mutations is synergistic with epistasis parameter $0.1$ and epistasis exponent $5$ (see ESM for epistasis function). Several similar plots with different sets of biological complexities are posted in the ESM. \\

\begin{figure}[h!]
  \begin{center}
	\vspace{1in}
    \includegraphics[width=0.55\textwidth]{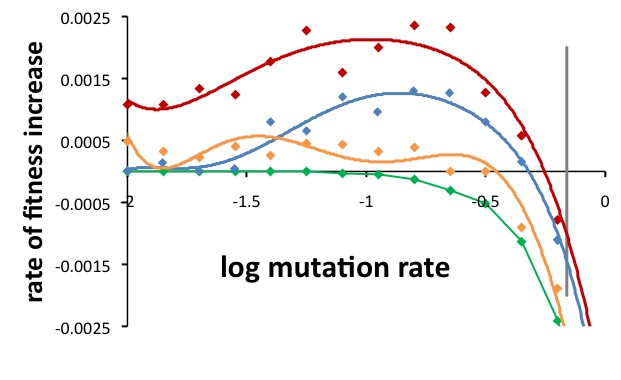}
  \end{center}
 \caption{}\label{fig:dwdt}
\end{figure}

\begin{figure}[h!]
    \begin{center}\includegraphics[width=0.6\textwidth]{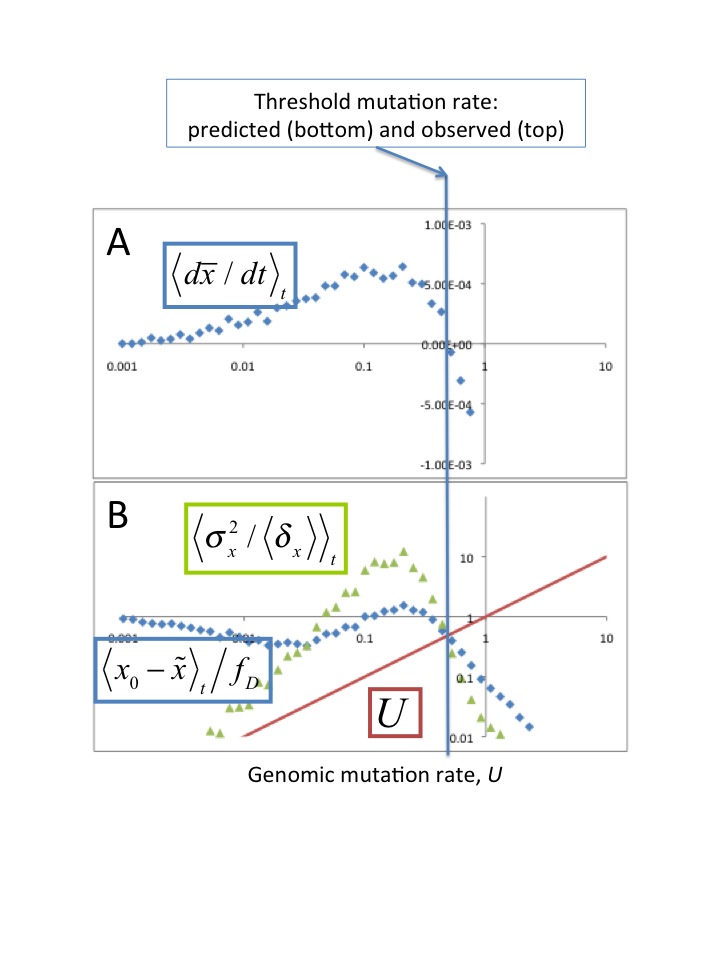}\end{center}
 \caption{}\label{fig:timeavg}
\end{figure}

\begin{figure*}[h!]
  \begin{center}
	\vspace{.5in}
    \includegraphics[width=0.95\textwidth]{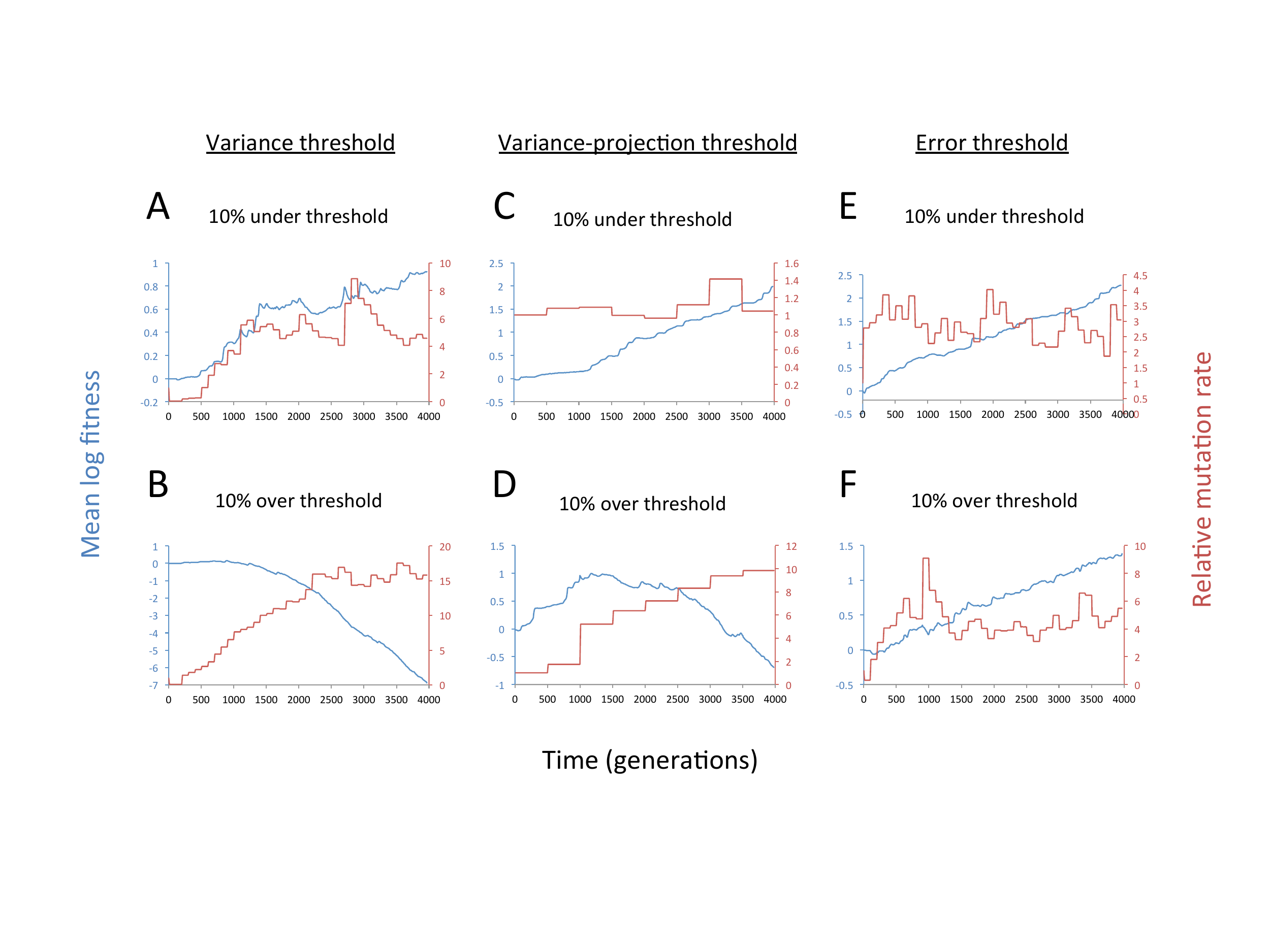}
  \end{center}
 \caption{}\label{fig:realtime}
\end{figure*}

\begin{figure}[h!]
  \centering
    \begin{center}\includegraphics[width=0.6\textwidth]{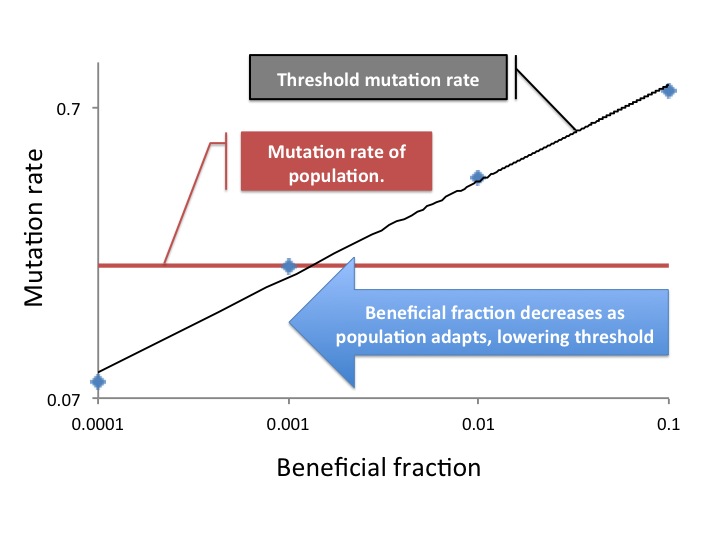}\end{center}
 \caption{}\label{fig:schematic}
\end{figure}

\noindent Figure 3. Real-time application of the different thresholds. Each panel plots a single representative simulation run. Simulations and parameters are the same as those in Fig. \ref{fig:dwdt}, with beneficial fraction set at $10^{-4}$. (\emph{A,B}) Variance threshold given by Eq. \eqref{condition1}. Every 100 generations, fitness variance is measured and $U$ is set equal to $-0.9 \sigma_x^2 / \left< \delta_x \right>$ (Panel \emph{A}) and $-1.1 \sigma_x^2 / \left< \delta_x \right>$ (Panel \emph{B}). (\emph{C,D}) Variance-projection threshold given by Eq. \eqref{fitness2}. Every 500 generations, fitness measurements are used to compute cumulants $\kappa_i(\tau)$ from which the $\kappa_2(\tau)$ are calculated (Appendix), and $U$ is set equal to $-0.9 \frac{1}{r} \sum\limits_{\tau=0}^{r-1} \kappa_2(\tau) / \left< \delta_x \right>$ (Panel \emph{C}) and  $-1.1 \frac{1}{r} \sum\limits_{\tau=0}^{r-1} \kappa_2(\tau) / \left< \delta_x \right>$ (Panel \emph{D}). (\emph{E,F}) Error threshold given by Eq. \eqref{ET}. Every 100 
generations, $U$ is set equal to $0.9 (x_0 - \tilde{x})/f_D$ (Panel \emph{E}) and $1.1 (x_0 - \tilde{x})/f_D$ (Panel \emph{F}). Our Criterion 1 appears to perform better for such real-time application than Criterion 2.\\

\noindent Figure 4. Schematic of how adaptive evolution and/or natural selection may be neutralized not as a result of increasing the mutation rate but as a result of a decreasing threshold mutation rate. The red line indicates the mutation rate of the population; the black line plots the threshold mutation rate as a function of the fraction of mutations that are beneficial (horizontal axis). The big blue arrow indicates that as a population adapts in a static environment, its supply of beneficial mutations is used up, resulting in a decreasing fraction of mutations that are beneficial. As this fraction decreases, the threshold mutation rate decreases, until eventually the threshold mutation rate is below the mutation rate of the population. If this relation is sustained, it seems reasonable to infer that extinction would ensue.\\

\newpage
\begin{table}[h] \caption{Summary of neutralizing conditions} %title of the table 
\centering \begin{tabular}{c ccccc} \hline\hline\\[-.8ex] Neutralizes:& Adaptive evolution & Natural selection\\ [-.2ex] & {\footnotesize (Criterion 1)} & {\footnotesize (Criterion 2)}\\ [0.8ex] \hline\hline \\[-.8ex]
  & $-U \left< \delta_x \right> > \sigma_x ^2$ &  $U f_D \gtrsim x_0-\bar{x}$  \\[-1ex] % Entering row contents 
\raisebox{1.5ex}{Sufficient:} & persistently & persistently$^*$  \\[2ex] Sufficient and & $- U \left< \delta_x \right> > \frac{1}{r} \sum\limits_{i = 0}^{r-1}\kappa_2 (i)$ \\[-1ex] % Entering row contents 
\raisebox{1.5ex}{\textit{somewhat} necessary:}& intermittently  \\[2ex]  & $-U \left< \delta_x \right> > \sigma_x ^2$ & $U f_D \ge x_0 - \tilde{x}$\\[-1ex] % Entering row contents 
\raisebox{1.5ex}{Sufficient and necessary:} & long-term average & long-term steady state  \\[2ex] % centering table % creating eight columns %inserting double-line 
\hline \end{tabular} \label{tab:hresult} \end{table}
\begin{centering}
$^*$ This condition holds when $L \gg N$

\end{centering}

\newpage

\newpage

\newpage

\newpage

\end{spacing}
 
\end{document}